\IfFileExists{revtex4-2.cls}{\newcommand{\CLASS}{revtex4-2}}{\IfFileExists{revtex4-1.cls}{\newcommand{\CLASS}{revtex4-1}}{	\newcommand{\CLASS}{revtex4}}}
\documentclass[prx,reprint,onecolumn,superscriptaddress,floatfix,aps,12pt]{\CLASS}

\usepackage{amssymb,amsmath,latexsym,bm,dsfont,graphicx,siunitx,physics}
\usepackage{makecell,newtx,xcolor}
\usepackage{braket}
\usepackage{mfirstuc}
\makeatletter\def\@pacs@name{DOI: }\makeatother

\makeatletter
\def\frontmatter@authorformat{\small}
\def\frontmatter@affiliationfont{\footnotesize\it}
\makeatother

\usepackage{caption}
\usepackage{ragged2e}
\usepackage{booktabs}
\usepackage{multirow}
\let\epsilon\varepsilon

\renewcommand{\tilde}{\widetilde}

\def\bibsection{\section*{\refname}}

\begin{document}
%%%%%%%%%%%%%%%%%%%%%%%%%%%%%%%%%%%%%%%%%%%%%%%%%%%%%%%%%%%%%%%%%%%%
\title{Observation of current-induced orbital quadrupole accumulation}
\author{Geun-Hee Lee}
%\email{ghlee0001@kaist.ac.kr}
\affiliation{Department of Physics, Korea Advanced Institute of Science and Technology (KAIST), Daejeon, Korea}
%\affiliation{Center for Quantum Dynamics of Angular Momentum, Pohang University of Science and Technology, Pohang, Korea}
\affiliation{Center for Quantum Dynamics of Angular Momentum, POSTECH, Pohang, Korea}

\author{Yubin Ji}
\affiliation{Department of Physics, Korea Advanced Institute of Science and Technology (KAIST), Daejeon, Korea}

\author{Yongho Park}
\affiliation{Department of Physics, Yonsei University, Seoul, Korea}
\affiliation{Center for Quantum Dynamics of Angular Momentum, POSTECH, Pohang, Korea}

\author{Changmin An}
\affiliation{Department of Physics, Korea Advanced Institute of Science and Technology (KAIST), Daejeon, Korea}
\affiliation{Center for Quantum Dynamics of Angular Momentum, POSTECH, Pohang, Korea}

\author{San Ko}
\affiliation{Department of Physics, Korea Advanced Institute of Science and Technology (KAIST), Daejeon, Korea}

\author{Hye-Won Ko}
\affiliation{Department of Physics, Korea Advanced Institute of Science and Technology (KAIST), Daejeon, Korea}
\affiliation{Department of Materials, ETH Zürich, Zürich, Switzerland}

\author{Jinseob Lim}
\affiliation{Department of Physics, Korea Advanced Institute of Science and Technology (KAIST), Daejeon, Korea}
\affiliation{Center for Quantum Dynamics of Angular Momentum, POSTECH, Pohang, Korea}

\author{Jung Hyun Oh}
\affiliation{Department of Physics, Korea Advanced Institute of Science and Technology (KAIST), Daejeon, Korea}
\affiliation{Center for Quantum Dynamics of Angular Momentum, POSTECH, Pohang, Korea}

\author{Farzad Mahfouzi}
\affiliation{Physical Measurement Laboratory, National Institute of Standards and Technology, Gaithersburg, Maryland, USA}
\affiliation{Department of Chemistry and Biochemistry, University of Maryland, College Park, USA}

\author{Byong-Guk Park}
\affiliation{Department of Materials Science and Engineering, KAIST, Daejeon, Korea}
\affiliation{Center for Quantum Dynamics of Angular Momentum, POSTECH, Pohang, Korea}

\author{Kab-Jin Kim}
\affiliation{Department of Physics, Korea Advanced Institute of Science and Technology (KAIST), Daejeon, Korea}

\author{Mark D. Stiles}
\affiliation{Physical Measurement Laboratory, National Institute of Standards and Technology, Gaithersburg, Maryland, USA}

\author{Kyoung-Whan Kim}
\affiliation{Department of Physics, Yonsei University, Seoul, Korea}
\affiliation{Center for Quantum Dynamics of Angular Momentum, POSTECH, Pohang, Korea}

\author{Paul M. Haney}
\affiliation{Physical Measurement Laboratory, National Institute of Standards and Technology, Gaithersburg, Maryland, USA}

\author{Kyung-Jin Lee$^\dagger$}
\email{kjlee@kaist.ac.kr}
\affiliation{Department of Physics, Korea Advanced Institute of Science and Technology (KAIST), Daejeon, Korea}
\affiliation{Center for Quantum Dynamics of Angular Momentum, POSTECH, Pohang, Korea}

%\date{\today}
%%%%%%%%%%%%%%%%%%%%%%%%%%%%%%%%%%%%%%%%%%%%%%%%%%%%%%%%%%%%%%%%%%%%%%%%%
%\vspace{10pt}
\begin{abstract}
\textbf{Spintronics~\cite{Hirohata2020} and
orbitronics~\cite{Bernevig2005,go2021,Cysne2025,Fukami2025} rely on
current-induced accumulations of magnetic dipoles:
spin~\cite{Kato2004,Stamm2017} and orbital angular
momentum~\cite{Choi2023,Lyalin2023,Marui2023}. However, electronic
orbitals inherently carry multipoles beyond the dipole, with the rank-2
orbital quadrupole as the leading term. Here we use polarization-resolved
Kerr microscopy to observe current-induced
orbital-quadrupole accumulation at the surfaces of Ti and Pt, metals with
markedly different spin--orbit-coupling strengths. By separating the
symmetric and antisymmetric components of the off-diagonal optical
conductivity, we isolate the time-reversal-even quadrupolar response from
the conventional time-reversal-odd magnetic-dipolar one, and find that the quadrupolar
optical response exceeds the dipolar one in both metals. First-principles analysis of the measured responses indicates that the
quadrupole accumulations are of the same order of magnitude in the two
metals despite their widely different spin--orbit-coupling strengths, consistent with a previously unidentified channel of
charge-to-orbital conversion that does not require spin--orbit coupling.
Our findings establish that current-induced orbital polarization is
fundamentally multipolar, expanding current-induced phenomena from the
dipolar to the multipolar regime and opening a route to electrical
control of orbital-ordered phases.}
\end{abstract}

%\pacs{\quad}

\maketitle

%%%%%%%%%%%%%%%%%%%%%%%%%%%%%%%%%%%%%%%%%%%%%%%%%%%%%%%%%%%%%%%%%%%%%%%%%%
%\vspace{10pt}
%Correspondence to: kjlee@kaist.ac.kr (K.-J. Lee$^\dagger$)
\newpage
%\section*{Main}
Multipolarity is intrinsic to orbital states and fundamentally distinguishes orbital physics from spin physics for conduction electrons.
An electron carries a spin-$1/2$ state, described by a $2\times2$
density matrix containing one charge and three spin components, with no
independent higher-rank multipoles. By contrast, an orbital state with
angular momentum quantum number $l$ is described by a
$(2l+1)\times(2l+1)$ density matrix: beyond one charge (rank-zero,
monopole) and three orbital angular momentum (OAM; rank-one, dipole)
components, the remaining $[(2l+1)^2-4]$ components constitute
higher-rank multipoles, led by the rank-two orbital quadrupole~\cite{Han2022}.
The orbital quadrupole, defined by the symmetrized product of OAM
operators $\{L_\mu,L_\nu\}$, is time-reversal even and is thus
symmetry-distinct from the time-reversal-odd spin and OAM. Importantly,
this higher-rank character does not imply a perturbatively smaller
response: within a fixed orbital manifold, the orbital quadrupole and
OAM map onto the electric quadrupole and the magnetic dipole,
respectively~\cite{Inui1990}, whose couplings enter the electromagnetic
multipole expansion at the same order~\cite{Landau1984}. The orbital
quadrupole must therefore be treated on an equal
footing with OAM.

The orbital quadrupole describes the shape and orientation of the
electronic wavefunction (Fig.~\ref{Fig:schematics}a; Supplementary
Note~1). It constitutes a leading order parameter in orbital-ordered phases, central to condensed-matter physics: in
strongly correlated systems, orbital order couples to spin, charge, and
lattice degrees of freedom, underlying metal--insulator transitions,
Kugel--Khomskii exchange, and orbital-selective
correlations~\cite{Tokura2000,Imada1998,kugel1982,Anisimov2002}. Many
hallmark equilibrium phenomena of orbital physics are governed
by orbital quadrupoles. Generating and controlling this quadrupolar
degree of freedom out of equilibrium---for instance, with an electric
current---would extend orbital physics beyond the equilibrium and dipolar regimes.

Generating nonequilibrium orbital quadrupoles does not require pre-existing real-space orbital
order. Even in crystals without orbital ordering, orbital hybridization
produces momentum-space orbital-quadrupole textures~\cite{Go2018} that
dynamically couple the OAM and orbital-quadrupole
sectors~\cite{Han2022} (Supplementary Note~2). Nonequilibrium orbital responses therefore generically engage the
orbital quadrupole alongside OAM, as shown theoretically for orbital pumping~\cite{Han2025}. Theories also predict that an electric
current generates orbital-quadrupole accumulation even in the absence of spin--orbit coupling (SOC): an
orbital-quadrupole Hall response has been predicted~\cite{Han2022}, and
we find that an intrinsic orbital-quadrupole
Edelstein effect generally emerges at surfaces (Supplementary Note~3), consistent with the
dc-field-induced reshaping of electronic wavefunctions~\cite{Mahfouzi2025_1}. However, direct experimental detection of
current-induced orbital-quadrupole accumulation has been lacking.

Here we close this gap by optically detecting current-induced
orbital-quadrupole accumulation at the surfaces of Ti and Pt, two metals
with markedly different SOC strengths. The central idea is that the time-reversal-even orbital quadrupole and the time-reversal-odd magnetic dipole contribute to distinct symmetry channels of the optical conductivity: symmetric and antisymmetric off-diagonal components, respectively. Exploiting this contrast, we isolate and quantify the quadrupolar optical response and find that it exceeds the dipolar one in both metals, establishing the orbital quadrupole as a substantial, previously unidentified channel of charge-to-orbital conversion that persists even in light elements.

\section*{Symmetry analysis of orbital-quadrupole generation and detection}
\begin{figure} % Do NOT use \begin{figure*}
	\centering
	\includegraphics[width=\textwidth]{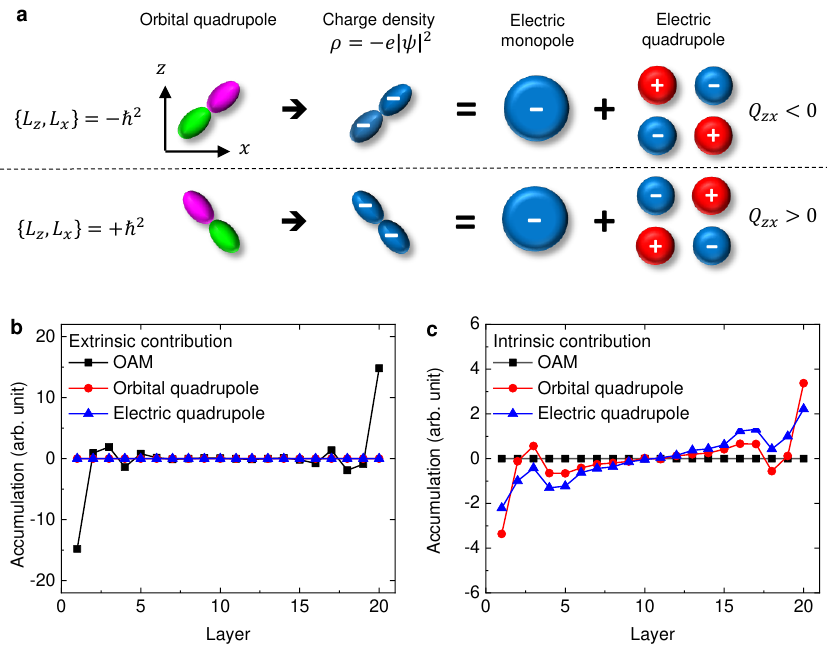} % for an image file named example_figure.*
	% Pick an appropriate width - in print, figures are usually one or two columns wide, which can
	% be approximated by 0.3\textwidth or 0.6\textwidth respectively. Use appropriate label sizes.        
    	% Captions go below figures
	\caption{\justifying \textbf{Current-induced orbital-quadrupole accumulation.} \textbf{a,} Correspondence between orbital quadrupole and electric quadrupole. The orbital quadrupole describes the directional anisotropy of an orbital state (magenta and green denote $+$ and $-$ signs of the wavefunction's phase). The anisotropic charge density of a state with nonzero $\{L_z,L_x\}$ decomposes into an electric monopole and an electric quadrupole $Q_{zx}$, whose sign tracks that of $\{L_z,L_x\}$ (top: $-\hbar^2$; bottom: $+\hbar^2$, where $\hbar$ is the reduced Planck constant).
\textbf{b,c,} First-principles calculations of current-induced accumulations of OAM, orbital quadrupole, and electric quadrupole in a 20-monolayer Ti slab from extrinsic (\textbf{b}) and intrinsic (\textbf{c}) mechanisms.}
\label{Fig:schematics}
\end{figure}

Because light couples to charge through its oscillating electric field,
an orbital quadrupole $\{L_\mu, L_\nu\}$ is detected optically through its mapped electric
quadrupole $Q_{\mu\nu}$---an anisotropic charge
distribution with $\mu\nu$ symmetry (Fig.~\ref{Fig:schematics}a).
The optical response of an accumulated orbital quadrupole is therefore
that of an electric quadrupole. 

%The characteristic charge distribution of quadrupoles requires specific broken symmetries. 
Theory predicts that a dc electric field along ${\hat{\mathbf{x}}}$ induces an accumulation of $\{L_z,L_x\}$, corresponding to $Q_{zx}$, at a surface normal to ${\hat{\mathbf{z}}}$~\cite{Han2022} (Supplementary Note~3). The symmetry requirement follows from the definition $Q_{\mu\nu} = \int r_\mu r_\nu \rho(\mathbf{r}) d^3\mathbf{r}$, where $\rho(\mathbf{r})$ is the charge density and $\mathbf{r}=(r_x,r_y,r_z)$: $Q_{zx}$ is even under both space inversion and time reversal, but odd under the mirror operations $\mathcal{M}_z$ and $\mathcal{M}_x$, whose mirror planes are perpendicular to ${\hat{\mathbf{z}}}$ and ${\hat{\mathbf{x}}}$, respectively. Both mirrors are broken in our geometry---$\mathcal{M}_z$ by the surface and $\mathcal{M}_x$ by the applied field---so surface accumulation of $Q_{zx}$ under an $x$-directed field is symmetry-allowed at a generic $z$-normal surface. We note that $\{L_z,L_x\}$ and $Q_{zx}$ transform identically under inversion, time reversal, and mirror operations and that a dc-field-induced change in the optical conductivity is also known as the electro-optic effect~\cite{Landau1984}.

To support this analysis, we compute current-induced accumulations of OAM, orbital quadrupole, and electric quadrupole in a 20-monolayer Ti slab (Methods). All three quantities are accumulated at surfaces (Figs.~\ref{Fig:schematics}b,c): the OAM accumulation arises from the extrinsic (Fermi-surface, scattering-dependent) mechanism, whereas both quadrupole accumulations arise from the intrinsic (Fermi-sea, scattering-independent) mechanism, consistent with their time-reversal symmetries.

We now turn to the optical detection scheme.
Magneto-optical Kerr effect (MOKE) measurements have been widely used to detect current-induced
spin~\cite{Kato2004,Stamm2017} and
OAM~\cite{Choi2023,Lyalin2023,Marui2023} accumulations, with a common assumption that the optical signals are entirely governed by the magnetic-dipole accumulations. As we demonstrate below, however, an electric-quadrupole accumulation also modifies the optical signals, through a distinct symmetry channel that allows it to be separated from the magnetic-dipole contribution (see Ref.~\cite{Mahfouzi2025_1} for a closely related theoretical analysis).

The relevant symmetry is encoded in the optical conductivity tensor. We decompose the off-diagonal optical conductivity into the symmetric part $\sigma_{\mu\nu}^{\rm S}=\left(\sigma_{\mu\nu}+\sigma_{\nu\mu}\right)/2$ and the antisymmetric part $\sigma_{\mu\nu}^{\rm A}=\left(\sigma_{\mu\nu}-\sigma_{\nu\mu}\right)/2$, with $\mu\ne\nu$.
By Onsager reciprocity, the symmetric part $\sigma_{\mu\nu}^{\rm S}$ is time-reversal even, whereas the antisymmetric part $\sigma_{\mu\nu}^{\rm A}$ is time-reversal odd~\cite{Landau1984}. This decomposition matches the symmetry of electric quadrupoles and magnetic dipoles: an electric quadrupole $Q_{\mu\nu}$ is symmetric and time-reversal even, whereas a magnetic dipole $M_\lambda$ can be represented by the antisymmetric, time-reversal-odd tensor $m_{\mu\nu}=\varepsilon_{\mu\nu\lambda}M_\lambda$ with the Levi-Civita symbol $\varepsilon_{\mu\nu\lambda}$. 
 
This symmetry correspondence dictates that $Q_{zx}$ maps to $\sigma^{\rm S}_{zx}$ and $M_y$ to $\sigma^{\rm A}_{zx}$, as we confirm within an orbital Rashba model (Supplementary Note~3). To leading order, the current-induced optical conductivity tensor $\boldsymbol{\sigma}$ takes the form
\begin{equation}
    \boldsymbol{\sigma} =  S
    \begin{bmatrix}
        0 & 0 & Q_{zx} \\
        0 & 0 & 0 \\
        Q_{zx} & 0 & 0
    \end{bmatrix} + T
    \begin{bmatrix}
        0 & 0 & M_{y} \\
        0 & 0 & 0 \\
        -M_{y} & 0 & 0
    \end{bmatrix},
    \label{Eq:opticalConductivity}
\end{equation}
where the rows and columns correspond to $x$, $y$, and $z$, and \(S\) and \(T\) are material-dependent conversion factors for the electric-quadrupole and magnetic-dipole responses, respectively. 

This tensor structure enables a straightforward experimental separation of the two contributions. In our geometry (Figs.~\ref{Fig:2}a,b), the incident and reflected beams lie in the $yz$ plane, so $s$-polarized light has its electric field along $\hat{\mathbf{x}}$, whereas $p$-polarized light has a component along $\hat{\mathbf{z}}$. The $s\rightarrow p$ cross-polarized response therefore probes the $zx$ component of $ \boldsymbol{\sigma}$, $\sigma_{zx}= S Q_{zx}-T M_y$, whereas the $p\rightarrow s$ response probes the $xz$ component, $\sigma_{xz}= S Q_{zx}+T M_y$ [Eq.~(\ref{Eq:opticalConductivity})]. The complex Kerr angles satisfy $\tilde{\theta}_{s\rightarrow p} \propto +\sigma_{zx}$ and $\tilde{\theta}_{p\rightarrow s} \propto -\sigma_{xz}$ as given by~\cite{Mahfouzi2025_1}
\begin{equation}
    \begin{split}
        \tilde{\theta}_{s\rightarrow p} &= \theta_{s \rightarrow p} + i \eta_{s \rightarrow p} = +\frac{n}{n^2-1} \mathcal{C}_{s\rightarrow p} \frac{\sigma_{zx}}{\sigma_{\rm{diag}}}, \\
        \tilde{\theta}_{p\rightarrow s} &= \theta_{p \rightarrow s} + i \eta_{p \rightarrow s} = -\frac{n}{n^2-1} \mathcal{C}_{p\rightarrow s} \frac{\sigma_{xz}}{\sigma_{\rm{diag}}},
    \end{split}
    \label{Eq:KerrFresnel}
\end{equation}
with the polarization rotation $\theta$, the ellipticity $\eta$, the refractive index $n$, the incidence-angle-dependent factor $\mathcal{C}_{s\rightarrow p (p\rightarrow s)}$, and the diagonal conductivity $\sigma_{\rm{diag}}$ (Methods). Consequently, $\tilde{\theta}_{s\rightarrow p}$ and $\tilde{\theta}_{p\rightarrow s}$ share the same sign for $M_y$ (Fig.~\ref{Fig:2}a) but have opposite signs for $Q_{zx}$ (Fig.~\ref{Fig:2}b). This sign contrast allows the two contributions to be disentangled.

\section*{Polarization-resolved current-induced reflection from Ti and Pt}

We deposit polycrystalline Ti films of varying thickness $t_{\rm Ti}$ and pattern them into wires of length 100~{\textmu}m along ${\hat{\mathbf{x}}}$ and width 20~{\textmu}m along ${\hat{\mathbf{y}}}$ (Methods). With a charge current applied along ${\hat{\mathbf{x}}}$, we measure $\tilde{\theta}_{s\rightarrow p}$ and $\tilde{\theta}_{p\rightarrow s}$ at an average incidence angle of $\phi_{\rm i}=20.6^\circ$ while scanning a laser beam along ${\hat{\mathbf{y}}}$ (Methods). 
Each Kerr angle is decomposed into the longitudinal MOKE (L-MOKE) component, $[\tilde{\theta}(+\phi_{\rm i})-\tilde{\theta}(-\phi_{\rm i})]/2$, which is sensitive to $M_y$ and $Q_{zx}$, and the polar MOKE (P-MOKE) component, $[\tilde{\theta}(+\phi_{\rm i})+\tilde{\theta}(-\phi_{\rm i})]/2$, which is sensitive to $M_z$ and $Q_{xy}$. 

Figures~\ref{Fig:2}c--\ref{Fig:2}f show line profiles of the L-MOKE and P-MOKE components of the current-induced changes in the polarization rotation $\theta$ and ellipticity $\eta$ for $s$- and $p$-polarized incident light. The L-MOKE components are spatially uniform across the wire ($|y|<10$~{\textmu}m). By contrast, the P-MOKE components are odd in $y$, following the symmetry of the current-induced Oersted field, which was previously identified as the dominant source of the polar signal~\cite{Choi2023}. We therefore focus on the L-MOKE signal in what follows.

\begin{figure} % Do NOT use \begin{figure*}
	\centering
	\includegraphics[width=\textwidth]{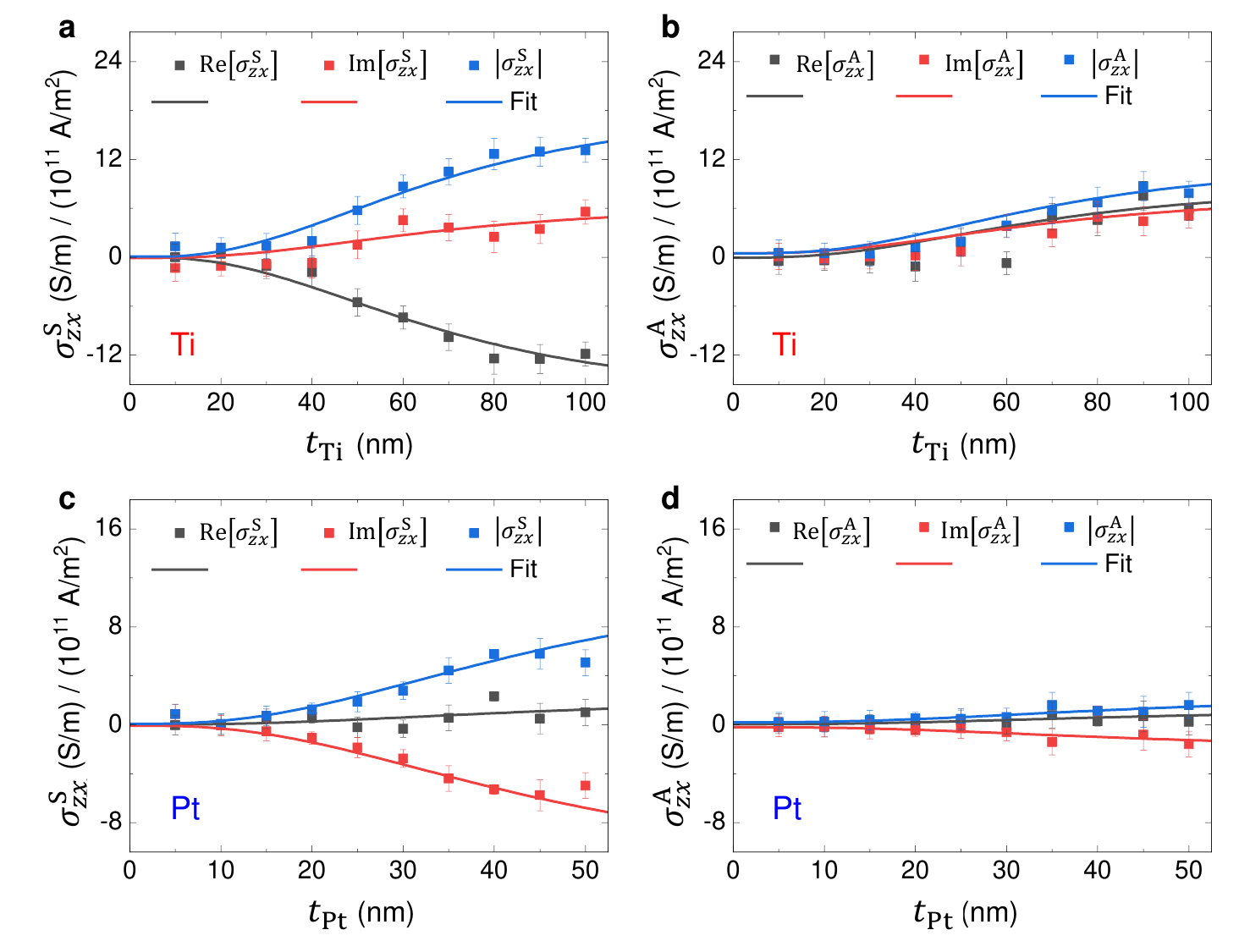} % for an image file named example_figure.*
	% Pick an appropriate width - in print, figures are usually one or two columns wide, which can
	% be approximated by 0.3\textwidth or 0.6\textwidth respectively. Use appropriate label sizes.
	% Captions go below figures
\caption{\justifying \textbf{Thickness dependence of the electric-quadrupole and magnetic-dipole responses in Ti (a,b) and Pt (c,d).} \textbf{a,b,} Current-induced symmetric optical conductivity $\sigma_{zx}^{\rm S}$ (\textbf{a}), associated with the electric quadrupole $Q_{zx}$, and antisymmetric optical conductivity $\sigma_{zx}^{\rm A}$ (\textbf{b}), associated with the magnetic dipole $M_y$, as a function of the Ti layer thickness $t_{\rm Ti}$. 
\textbf{c,d,} $\sigma_{zx}^{\rm S}$ (\textbf{c}) and $\sigma_{zx}^{\rm A}$ (\textbf{d}) as a function of the Pt layer thickness $t_{\rm Pt}$. 
%Corresponding modified complex Kerr angles $\tilde{\theta}_Q$ for $Q_{zx}$ and $\tilde{\theta}_M$ for $M_y$, defined in Eq.~(\ref{Eq:newKerrangle}). 
The real and imaginary parts, and the magnitude are shown in each panel. Solid lines are least-squares fits to an electro-optic profile convolved with the evanescent light [Eq.~(\ref{Eq:diffusionFit})]. Error bars represent 95\% confidence intervals obtained from line-scan averaging at each thickness.}
\label{Fig:3}
\end{figure}

At each current, we fit the plateau of the L-MOKE component over $|y|<10$~{\textmu}m and plot the fitted values as a function of the current density $j$ in Figs.~\ref{Fig:2}g--\ref{Fig:2}j. For both incident polarizations, $\theta$ and $\eta$ vary linearly with the current density, with negligible current-even components. This odd-in-current response rules out Joule heating ($\propto j^{2}$) as the dominant origin of the signals and establishes that the off-diagonal optical conductivities $\sigma_{zx}$ and $\sigma_{xz}$ are induced linearly by the applied current density. %Importantly, the $s\to p$ and $p\to s$ signals differ markedly. Because the magnetic dipole $M_y$ enters the off-diagonal conductivity antisymmetrically ($\sigma_{xz}=-\sigma_{zx}$) whereas the electric quadrupole $Q_{zx}$ enters symmetrically ($\sigma_{xz}=\sigma_{zx}$), a purely magnetic-dipolar response would fix the ratio of the two signals through the Fresnel factors alone [Eq.~(\ref{Eq:KerrFresnel})]; the observed deviation from this ratio reveals a substantial electric-quadrupole contribution. This contrasts sharply with previous magneto-optical studies~\cite{Kato2004,Stamm2017,Choi2023,Lyalin2023,Marui2023}, in which the current-induced signals were attributed entirely to magnetic-dipole (spin or OAM) accumulation.
We then convert the measured complex Kerr angles into the symmetric and antisymmetric conductivity components, associated with the electric-quadrupole and magnetic-dipole contributions, using Eq.~(\ref{Eq:KerrFresnel}) (Methods). Figures~\ref{Fig:3}a,b show $\sigma_{zx}^{\rm S}$ and $\sigma_{zx}^{\rm A}$ as a function of Ti thickness. Across the full thickness range, the symmetric component is larger in magnitude: the electric-quadrupole contribution is the larger of the two in Ti.

We fit the thickness dependence of $\sigma_{zx}^{\rm S}$ and $\sigma_{zx}^{\rm A}$ with a model in which the electro-optic response is generated at the top and bottom surfaces of the film with equal magnitude and opposite sign, and decays exponentially away from each surface over a length scale $\ell$. Weighting this profile with the evanescent light field yields~\cite{Stamm2017,Lyalin2023}:
\begin{equation}
\sigma_{zx}^{\rm S (A)} = \sigma_{{\rm sat}}^{\rm S (A)}\frac{\exp\left(\frac{t}{2\ell}\right)}{\cosh\left(\frac{t}{2\ell}\right)}\left(\frac{\left(\exp\left[-t\left(\frac{1}{\ell_{\rm P}}-\frac{1}{\ell}\right)\right]-1\right)\exp\left(-\frac{t}{\ell}\right)}{1-\frac{\ell_{\rm P}}{\ell}}-\frac{\exp\left[-t\left(\frac{1}{\ell_{\rm P}}+\frac{1}{\ell}\right)\right]-1}{1+\frac{\ell_{\rm P}}{\ell}}\right), 
\label{Eq:diffusionFit}
\end{equation}
where $\sigma_{{\rm sat}}^{\rm S (A)}$ is the saturation value, $t$ is the film thickness, and $\ell_{\rm P}$ is the penetration depth of the light calculated from $n$ ($\ell_{\rm P}=31.6$~nm for Ti; Methods). 
We fit both channels with a common $\ell$, as the fit uncertainties do not justify channel-specific values. The fits yield $\sigma_{{\rm sat}}^{\rm S}=[(-16.17+6.05i)\pm1.77]\times10^{-11}$~(S/m)$\cdot$(m$^2$/A), $\sigma_{{\rm sat}}^{\rm A}=[(8.28+6.61i)\pm2.00]\times10^{-11}$~(S/m)$\cdot$(m$^2$/A), and $\ell=(24\pm8)$~nm, where the uncertainties denote 95\% confidence intervals. In magnitude, $\sigma_{\rm sat}^{\rm S}$ exceeds $\sigma_{\rm sat}^{\rm A}$ by a factor of $\sim$1.6, showing that the electric-quadrupole contribution to the optical conductivity exceeds the magnetic-dipole one. We note that $\ell$ should not be interpreted as a real-space accumulation length: first-principles calculations~\cite{Mahfouzi2025_2} show that the orbital accumulations are highly localized near the surfaces, while the length extracted from the thickness dependence of
the optical signal is much longer. We therefore regard $\ell$ as an effective length characterizing the thickness dependence of the electro-optic response.

To investigate the role of SOC in the quadrupolar response, we repeat the polarization-resolved measurements on polycrystalline Pt films, in which strong SOC underpins the spin Hall effect~\cite{Stamm2017}. Figures~\ref{Fig:3}c,d show the symmetric and antisymmetric current-induced optical conductivities as a function of Pt thickness $t_{\rm Pt}$. The fits with Eq.~(\ref{Eq:diffusionFit}) and $\ell_{\rm P}=23.9$~nm for Pt (Methods) yield $\sigma_{{\rm sat}}^{\rm S}=[(1.99-10.94i)\pm0.69]\times10^{-11}$~(S/m)$\cdot$(m$^2$/A), $\sigma_{{\rm sat}}^{\rm A}=[(1.15-1.72i)\pm0.60]\times10^{-11}$~(S/m)$\cdot$(m$^2$/A), and $\ell=(14\pm3)$~nm. The symmetric (electric-quadrupole) component exceeds the antisymmetric (magnetic-dipole) one by a factor of $|\sigma^{\rm S}_{\rm sat}|/|\sigma^{\rm A}_{\rm sat}|\approx5$, an even larger ratio than in Ti ($\sim$1.6). This dominance of the symmetric channel over the antisymmetric channel is consistent with theoretical calculations~\cite{Mahfouzi2025_1}.
Thus, even in the archetypal spin Hall metal, the dominant current-induced modification of the optical response is quadrupolar, not dipolar. This finding bears directly on the interpretation of previous current-induced magneto-optical experiments using only a single incident polarization~\cite{Kato2004,Stamm2017,Choi2023,Lyalin2023,Marui2023}: such a measurement generally mixes $\sigma_{zx}^{\rm S}$ and $\sigma_{zx}^{\rm A}$, and therefore, cannot unambiguously distinguish spin/OAM from quadrupolar contributions.

Finally, we infer the microscopic multipole accumulations from $\sigma_{{\rm sat}}^{\rm S}$ and $\sigma_{{\rm sat}}^{\rm A}$, using first-principles conversion factors between each induced multipole density and its associated optical conductivity (Methods; Table~\ref{tab:accumulation}). The orbital-quadrupole and electric-quadrupole accumulations in Pt exceed those in Ti by factors of only 4.8 and 1.8, respectively. Despite the much stronger SOC of Pt, both quadrupolar accumulations remain of the same order in the two metals. Together with theories predicting finite quadrupole generation without SOC (Ref.~\cite{Han2022} and Supplementary Note~3), this result indicates that strong SOC is not required and that the response can remain effective in light 3$d$ metals.

For the antisymmetric magnetic-dipole channel, the optical conductivity alone does not determine the relative weights of spin and OAM; the spin and OAM values in Table~\ref{tab:accumulation} are therefore single-channel estimates, each obtained by attributing the full signal to one channel. These estimates can nevertheless be assessed against the known SOC strengths. A spin-only interpretation for Ti yields a value approximately two orders of magnitude larger than the corresponding Pt value. Given the weak SOC of Ti, such a value is implausible, indicating an OAM-dominated response in the light element Ti, consistent with Ref.~\cite{Choi2023}. Moreover, the spin accumulation in Pt~\cite{Stamm2017} and the OAM accumulation in Ti~\cite{Choi2023} reported in previous single-polarization MOKE studies exceed our corresponding estimates by factors of 2.2 and 2.6, respectively. This difference is consistent with a quadrupolar contribution that is ignored in the single-polarization signals analyzed in those studies, underscoring the need for polarization-resolved detection.

\section*{Discussion and outlook}
Current-induced orbital-quadrupole accumulation extends nonequilibrium orbital physics beyond the dipolar sector, with two direct implications. 
First, orbital torques call for reinterpretation. Current-driven torques in light elements have been attributed predominantly to OAM accumulation~\cite{Ding2020,Kim2021,Sala2022,Hayashi2023,Krishnia2023,Zheng2024}. However, theories suggest that torques also have orbital-quadrupole contributions: quadrupole pumping by magnetization dynamics~\cite{Han2025} implies, by Onsager reciprocity, a reciprocal quadrupolar torque, and orbital-quadrupole exchange~\cite{kugel1982} yields a quadrupolar torque on spin--orbit-coupled ferromagnets~\cite{Lee2026}. Our measurements show that the orbital-quadrupole accumulation is not negligible, motivating investigation of its contribution to current-induced torques. This quadrupolar channel may also bear on the puzzle of orbital relaxation lengths~\cite{Guan2026,Urazhdin2026}---long in several experiments~\cite{Sala2022,Choi2023,Hayashi2023,Gao2025} yet predicted to be very short for OAM~\cite{Belashchenko2023,Urazhdin2023,Rang2024,Mahfouzi2025_2}. 
Whether the orbital quadrupole, which has different symmetries from OAM, explains the observations remains an open question.
%Because the orbital quadrupole and OAM have different symmetries, they may have different relaxation lengths; whether this distinction explains the observations remains an open question.
 
Second, our results supply the missing ingredient for electrical control of orbital order: electrical control of magnetic order rests on two ingredients---current-induced spin accumulation and spin--spin exchange. Orbital-order control requires the analogous pair---current-induced orbital-quadrupole accumulation and quadrupole--quadrupole coupling. Our work establishes the former, while the latter is embodied in Kugel--Khomskii exchange~\cite{kugel1982}. Because the conjugate fields of a quadrupole---electric-field gradients or strain---are difficult to generate and switch in devices, a uniform current offers a practical route to manipulating orbital order, including multipolar order parameters in altermagnets~\cite{Bhowal2024,Mcclarty2024,Hayami2024}, and exciting orbitons~\cite{Saitoh2001}.
 
Beyond these implications, the orbital and electric quadrupoles share the mirror, inversion, and time-reversal symmetries of the quantum metric---the real part of the quantum geometric tensor~\cite{Provost1980}. Moreover, the electric-quadrupole moment of Bloch states has been linked theoretically to the quantum metric~\cite{Lapa2019,Gao2019,Daido2020}. Whether current-induced quadrupole accumulation can thus serve as a nonequilibrium probe of Bloch-state quantum geometry, complementing nonlinear-transport and spectroscopic approaches~\cite{Gao2023,Wang2023,Tian2023,Kim2025,Sala2025}, is an intriguing open question. These directions position current-induced orbital-quadrupole accumulation as a versatile platform for fundamental studies and applications.

\newpage
\def\bibsection{\section*{Main references}}

\newpage

\clearpage

\begin{figure} % Do NOT use \begin{figure*}
\centering
\includegraphics[width=\textwidth]{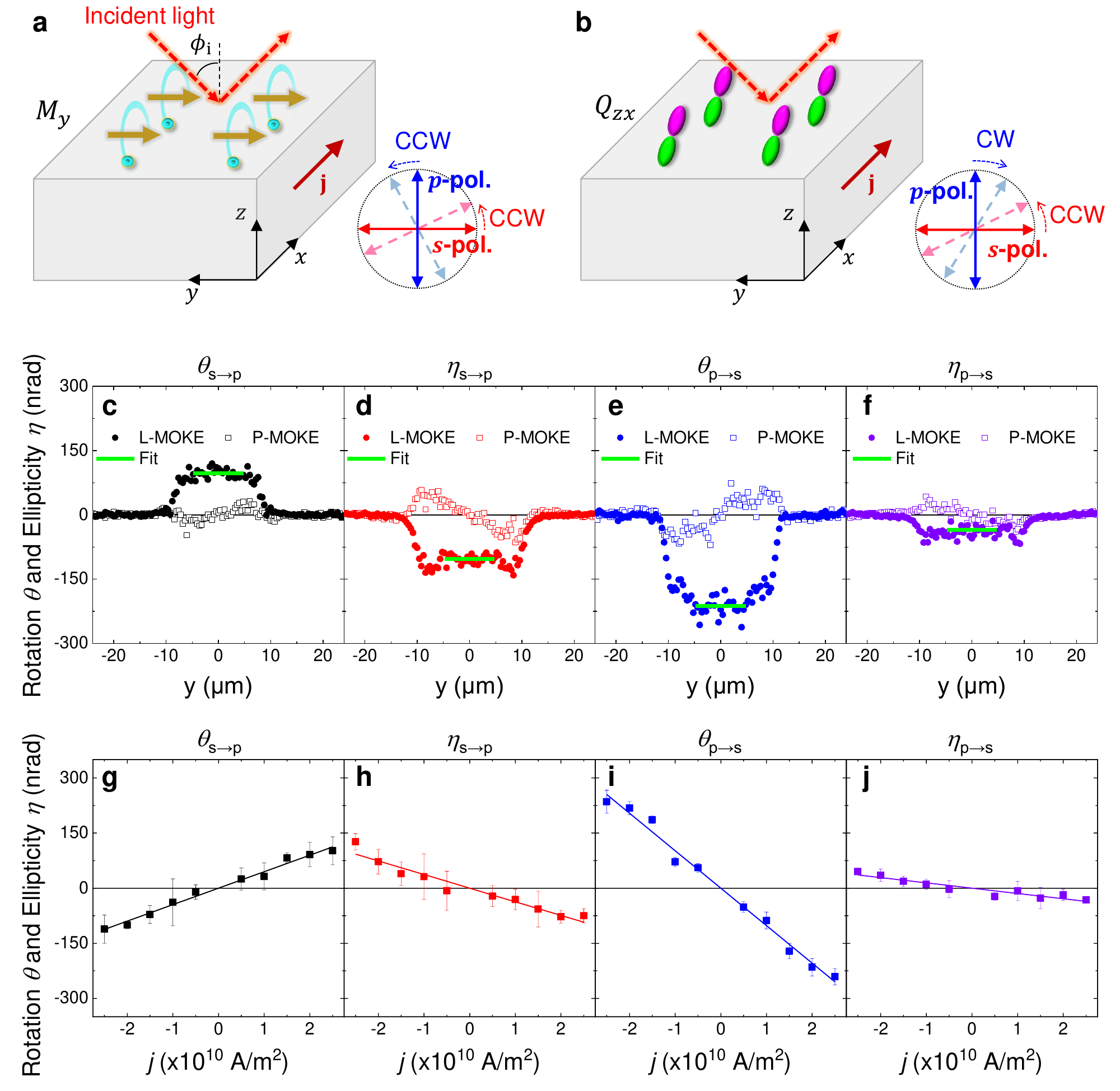} % for an image file named example_figure.*
% Pick an appropriate width - in print, figures are usually one or two columns wide, which can
% be approximated by 0.3\textwidth or 0.6\textwidth respectively. Use appropriate label sizes.
\caption{\justifying \textbf{Polarization-resolved optical detection of current-induced electric-quadrupole and magnetic-dipole accumulations.} \textbf{a,b,} Optical detection geometries with light incident at an average angle $\phi_{\rm i}$. An in-plane current density $j$ generates a magnetic dipole $M_y$ assciated with spin and OAM accumulations (\textbf{a}), and an electric quadrupole $Q_{zx}$ from orbital-quadrupole accumulation (\textbf{b}). $M_y$ rotates $s$- and $p$-polarized light in the same sense (both counterclockwise, CCW), whereas $Q_{zx}$ rotates them oppositely (CCW for $s$; clockwise, CW, for $p$).}
\label{Fig:2}
\end{figure}

\begin{figure} % Do NOT use \begin{figure*}
\ContinuedFloat
\caption{\justifying \textbf{Polarization-resolved optical detection of current-induced electric-quadrupole and magnetic-dipole accumulations (continued).} \textbf{c--f,} Spatial profiles of the current-induced Kerr rotation $\theta$ and ellipticity $\eta$ across the Ti wire as a function of probe position $y$: $\theta_{s\to p}$ (\textbf{c}), $\eta_{s\to p}$ (\textbf{d}), $\eta_{p\to s}$ (\textbf{e}), and $\theta_{p\to s}$ (\textbf{f}). Each panel shows the longitudinal (L-MOKE) and polar (P-MOKE) components. Solid green lines are fits to the longitudinal plateau over $-10$~{\textmu}m $\le y \le$ $10$~{\textmu}m. The data were obtained from a device with $t_{\rm Ti}=100$ nm under a current density of $2.5\times10^{10}$ A/m$^2$. \textbf{g--j,} The four longitudinal optical channels, $\theta_{s\to p}$ (\textbf{g}), $\eta_{s\to p}$ (\textbf{h}), $\eta_{p\to s}$ (\textbf{i}), and $\theta_{p\to s}$ (\textbf{j}), as a function of applied current density $j$. Lines are linear fits. Error bars represent 95\% confidence intervals obtained from line-scan averaging of the plateau at each current.}
\end{figure}

\clearpage

\clearpage
%\newpage

\begin{table}[htbp]
\centering
\setlength{\tabcolsep}{8pt}
\caption{\justifying \textbf{Estimates of accumulations required to account for the measured
current-induced optical conductivities.} Orbital- and electric-quadrupole
values are obtained from the symmetric conductivity $\sigma^{\rm S}_{zx}$,
and magnetic-dipole (spin, OAM) values from the antisymmetric conductivity
$\sigma^{\rm A}_{zx}$. The orbital- and electric-quadrupole entries are separate calibrations of the same symmetric quadrupolar channel and should not be interpreted as independently resolved additive contributions. The spin and OAM values are also single-channel estimates, each obtained by attributing the entire antisymmetric signal to that channel. For comparison, the last column lists previously reported
values: the OAM accumulation in Ti from Ref.~\cite{Choi2023} and the spin
accumulation in Pt from Ref.~\cite{Stamm2017}. $\mu_{\rm B}$ is the Bohr
magneton; $Q_{\rm unit}=eac/4$ for Ti and $Q_{\rm unit}=eb^2/4$ for Pt,
where $e$ is the elementary charge, $a$ and $c$ are the in-plane and
out-of-plane lattice constants of hcp Ti, and $b$ is the lattice constant
of fcc Pt.}
\label{tab:accumulation}
\begin{tabular}{llccc}
\toprule
\multicolumn{2}{c}{Accumulation} & Ti & Pt & Literature \\
\multicolumn{2}{c}{(per atom per $10^{11}$~A m$^{-2}$)} \\
\midrule
\multicolumn{2}{l}{Orbital quadrupole ($\hbar^2$)} & $(3.76\pm0.26)\times10^{-5}$ & $(1.82\pm0.11)\times10^{-4}$ & --- \\
\multicolumn{2}{l}{Electric quadrupole ($Q_{\rm unit}$)} & $(5.18\pm0.53)\times10^{-6}$ & $(9.46\pm0.59)\times10^{-6}$ & --- \\
\midrule
\multirow{2}{*}{Magnetic dipole}
& Spin ($\mu_{\rm B}$) & $(2.00\pm0.38)\times10^{-3}$ & $(2.32\pm0.67)\times10^{-5}$ & $5.0\times10^{-5}$ (Pt~\cite{Stamm2017}) \\
& OAM ($\mu_{\rm B}$)  & $(1.37\pm0.26)\times10^{-5}$ & $(7.74\pm2.25)\times10^{-6}$ & $3.5\times10^{-5}$ (Ti~\cite{Choi2023}) \\
\bottomrule
\end{tabular}
\end{table}

\clearpage
\section*{Methods}
\subsection*{Density functional theory calculation of multipole accumulation}
First-principles density functional theory calculations were performed using the OpenMX package~\cite{Ozaki2003}, which employs norm-conserving pseudopotentials~\cite{Bachelet1982} and pseudo-atomic localized basis functions~\cite{Ozaki2003}. The exchange-correlation functional was treated within the generalized gradient approximation~\cite{Perdew1996}. A periodic slab supercell consisting of 20 monolayers of hexagonal close-packed Ti was constructed with lattice constants of $a=0.295$~nm and $c=0.468$~nm, and a vacuum layer of approximately 2.5~nm was introduced along the surface-normal direction to avoid spurious interactions between periodic images. The pseudo-atomic orbital (PAO) basis set \(s2p2d2\) was used, and the self-consistent calculations were carried out with a \(\boldsymbol{k}\)-point mesh of \(28 \times 16 \times 1\). Once the ground state was converged, the layer-resolved accumulation was evaluated in post-processing using a Green's-function formalism within the Kubo linear-response framework. Intrinsic and extrinsic contributions were separated based on symmetry considerations~\cite{Bonbien2020}. Convergence of the accumulation was verified with a denser \(\boldsymbol{k}\)-point mesh of \(114 \times 114 \times 1\), and a level broadening of \(37~\text{meV}\) was employed to reproduce the experimental longitudinal electrical conductivity.

OAM operators were represented as \(18\times18\) matrices in the \(s2p2d2\) PAO basis within the atom-center approximation. Using the three OAM components \((L_x,L_y,L_z)\), the orbital quadrupole operators were defined as the symmetrized products \(\{ L_i,L_j\}\) with \(i,j\in\{ x,y,z\}\)~\cite{Han2022}.

The quantum-mechanical electric quadrupole operator for a single electron within the atom-center approximation is defined as~\cite{Kusunose2008,Hayami2018PRB,Kusunose2020,Hayami2018JPSJ}
\[
Q_2^m=-e\sqrt{\frac{4\pi}{5}}r_a^2Y_2^m(\hat{\mathbf{r}}_a).
\]
Here, \(Y_2^m\) denotes the spherical harmonic with \(-2\le m\le 2\), and \(\mathbf{r}_a=\mathbf{r}-\mathbf{R}_a\) is the electron position measured from the atomic site \(\mathbf{R}_a\). A more familiar representation of the electric quadrupole operator is obtained using cubic harmonics:
\[\begin{gathered}
Q_u \equiv Q_{3z^2-r^2} \equiv Q_2^0, \qquad
Q_v \equiv Q_{x^2-y^2} \equiv \frac{1}{\sqrt{2}}\left(Q_2^{-2}+Q_2^{2}\right),\\
Q_{xx} \equiv -\frac{1}{3}\left(er_a^2+Q_u\right)+\frac{1}{\sqrt{3}}Q_v, \qquad
Q_{yy} \equiv -\frac{1}{3}\left(er_a^2+Q_u\right)-\frac{1}{\sqrt{3}}Q_v, \qquad
Q_{zz} \equiv -\frac{1}{3}\left(er_a^2-2Q_u\right),\\
Q_{xy} \equiv \frac{i}{\sqrt{6}}\left(Q_2^{-2}-Q_2^{2}\right), \qquad
Q_{yz} \equiv \frac{i}{\sqrt{6}}\left(Q_2^{-1}+Q_2^{1}\right), \qquad
Q_{zx} \equiv \frac{1}{\sqrt{6}}\left(Q_2^{-1}-Q_2^{1}\right).
\end{gathered}\]
The Bloch sum of PAOs is written as
\[
|\Phi_{\alpha\mathbf k}\rangle=\frac{1}{\sqrt{N}}\sum_{a}e^{i\mathbf k\cdot\mathbf{R}_a}|\phi_{\alpha,\mathbf{R}_a}\rangle ,
\]
where \(|\phi_{\alpha,\mathbf{R}_a}\rangle=|R_{nl,\mathbf{R}_a}\rangle|\tilde{\phi}_{lm,\mathbf{R}_a}\rangle\) is a PAO consisting of the radial part \(|R_{nl,\mathbf{R}_a}\rangle\) and the angular part \(|\tilde{\phi}_{lm,\mathbf{R}_a}\rangle\). Here, \(\alpha=(n,l,m)\), where \(n\) labels the radial part for a given \(l\), \(l\) is the orbital angular momentum quantum number, and \(m\) labels the real component for the given \(l\). In this basis, the matrix element of the electric quadrupole operator is expressed as
\[
\left[Q_i\right]_{\alpha\beta}(\mathbf k)
=\langle\Phi_{\alpha\mathbf k}|Q_i|\Phi_{\beta\mathbf k}\rangle
=\frac{1}{N}\sum_{a,b}e^{i\mathbf k\cdot(\mathbf{R}_a-\mathbf{R}_b)}
\langle \phi_{\alpha,\mathbf{R}_b}|Q_i|\phi_{\beta,\mathbf{R}_a}\rangle
\]
\[
=\sum_{\Delta\mathbf R}e^{i\mathbf k\cdot\Delta\mathbf R}
\langle \phi_{\alpha,0}|Q_i|\phi_{\beta,\Delta\mathbf R}\rangle ,
\]
where \(\Delta\mathbf{R}=\mathbf{R}_a-\mathbf{R}_b\).

To isolate the on-site contribution, we further approximate the matrix element by inserting the PAO projection operator \(\sum_{\gamma}|\phi_{\gamma,0}\rangle\langle\phi_{\gamma,0}|\) where the PAOs at the same atomic site are orthonormal. This gives
\[
\left[Q_i\right]_{\alpha\beta}(\mathbf k)
\approx\sum_{\Delta\mathbf R}e^{i\mathbf k\cdot\Delta\mathbf R}\sum_{\gamma}
\langle \phi_{\alpha,0}|Q_i|\phi_{\gamma,0}\rangle\langle\phi_{\gamma,0}|\phi_{\beta,\Delta\mathbf R}\rangle .
\]
For the on-site matrix elements, the electric quadrupole operator can be separated into radial and angular parts:
\[
\langle \phi_{\alpha,0}|Q_i|\phi_{\gamma,0}\rangle
=\langle R_{nl,0}|r^2_0|R_{n'l',0}\rangle\langle\tilde{\phi}_{lm,0}|\tilde{Q}_i|\tilde{\phi}_{l'm',0}\rangle,
\]
where \(\tilde{Q}_i\) denotes the angular part of the electric quadrupole operator. 

The angular matrix element \(\langle\tilde{\phi}_{lm,0}|\tilde{Q}_i|\tilde{\phi}_{l'm',0}\rangle\) can be evaluated using the Wigner--Eckart theorem~\cite{Inui1990}.  In the present work, the corresponding multipole matrix elements are adopted from the Supplemental Material of Ref.~\cite{Hayami2018JPSJ}. The radial matrix element is calculated from the PAO radial functions as
\[
\langle R_{nl,0}|r^2_0|R_{n'l',0}\rangle
=\int_0^{r_{cut}}dr_0 R_{nl}^*(r_0)r_0^4R_{n'l'}(r_0),
\]
where \(r_{cut}\) is the cutoff radius of the PAO radial function~\cite{Ozaki2003}.

The factor \(\langle\phi_{\gamma,0}|\phi_{\beta,\Delta\mathbf R}\rangle\) is the PAO overlap integral, which measures the spatial overlap between two PAOs centered at different atomic sites. Through this overlap factor, the local on-site electric-quadrupole matrix element is extended to include inter-site contributions in an approximate manner. Therefore, the electric quadrupole operator is evaluated within the atom-center approximation together with the PAO basis projection approximation. In this scheme, the dominant on-site contribution is treated explicitly, while the inter-site contribution is incorporated through the PAO overlap integral.

\subsection*{Device fabrication}
Ti thin films were deposited on $c$-cut $\rm{Al}_2 \rm{O}_3$ substrates by dc magnetron sputtering under the following conditions: a base pressure of $1.3 \times 10^{-6}$~Pa, a working Ar pressure of 0.4~Pa, and an Ar flow rate of $3.4 \times 10^{-2}$~Pa$\cdot$m$^3$/s. To prevent natural oxidation of the Ti layer, capping layers of 2~nm MgO and 2~nm Ta were subsequently deposited. Pt thin films were deposited under the same conditions.
%The resistivity of Ti thin films remains approximately $3~\mu\Omega \cdot \mathrm{m}$ across the entire range of \(t_{\rm{Ti}}\) studied (Extended Data Fig.~\ref{Fig:S5}). %The slight enhancement at small thicknesses is attributed to interfacial scattering.

X-ray diffraction (XRD) measurements confirmed that the Ti thin films have a hexagonal close-packed (hcp) structure (Extended Data Fig.~\ref{Fig:S1}). %Two $2\theta$ peaks were observed at positions consistent with the $(0002)$ and $(10\bar{1}0)$ planes of hcp Ti. %($38.7^\circ$, $40.4^\circ$ for $c=0.468$ nm and $\lambda_{\mathrm{X-ray}}=0.154$ nm). 
On the other hand, XRD measurements confirmed that the Pt thin films have a face-centered cubic (fcc) structure (Extended Data Fig.~\ref{Fig:SPt}).

After deposition, the films were patterned into wires with dimensions of 100~{\textmu}m$\times$20~{\textmu}m using conventional photolithography and Ar ion etching. Ti (5~nm)/Au (100~nm) electrode pads were then fabricated by photolithography and sputtering to establish electrical connections with the measurement instruments. An optical image of the final Ti devices is shown in Extended Data Fig.~\ref{Fig:S2}.

\subsection*{Optical setup}

Current-induced Kerr responses on the Ti and Pt surfaces were measured using a focused-laser scanning system, schematically illustrated in Extended Data Fig.~\ref{Fig:S3}. All measurements were conducted at room temperature.

A continuous-wave laser at 660~nm first passes through a polarizer to produce $s$-polarized light, and then traverses the first beam splitter. The incidence angle, defined as $\phi_{\rm i} = \tan^{-1}(\Delta_{\rm{BS}}/f)$, is controlled by the $y$-directional displacement of the first beam splitter ($\Delta_{\rm{BS}}$), where $f=4\;\rm{mm}$ is the effective focal length of the objective lens. Before reaching the objective lens, the light passes through a half-wave plate (HWP1) that sets the incident polarization: when the fast axis of HWP1 is parallel to the $s$-polarization, the light remains $s$-polarized, whereas rotating the fast axis by $\pi/4$ converts the polarization to $p$. The light is then focused onto the wire using a $\times$50 objective lens (numerical aperture 0.55), yielding a beam size of 1.5~{\textmu}m and an optical power of 3.5~mW. These parameters were maintained identical for all measurements. 

An AC current at 1013~Hz was applied to the devices for current-induced polarimetry, and the first-harmonic change in light polarization at the AC frequency was extracted using a lock-in amplifier. The negative polarity of the AC current shown in Fig.~\ref{Fig:2} denotes a reversed current direction.

After reflection from the sample, the light is analyzed using two photo-detectors. A second beam splitter directs one beam to a reference photo-detector that measures the intensity of the reflected light ($I_0$), while the other beam passes through a second half-wave plate (HWP2) and a Wollaston prism (W.P.) before being detected by a balanced photo-detector ($\Delta I$). The fast axis of HWP2 is fixed at $\pi/8$ to balance the photo-detector signals in the absence of an applied current. 

By adjusting the angle of HWP1, both $\tilde{\theta}_{s \rightarrow p}$ and $\tilde{\theta}_{p \rightarrow s}$ can be measured. The Jones vector immediately before the balanced photo-detector is given by:
\begin{equation*}
    \begin{bmatrix}
    E^{(2)}_s \\
    E^{(2)}_p
    \end{bmatrix}
    =M_{\rm{HWP2}}M_{\rm{HWP1}}M_{\rm{refl}}M_{\rm{HWP1}}
    \begin{bmatrix}
    E^{(1)}_s \\
    E^{(1)}_p
    \end{bmatrix},
\end{equation*}
where $E^{(1)}_{s,p}$ and $E^{(2)}_{s,p}$ denote the $s$- and $p$-components of the electric field after the first beam splitter and before the balanced photo-detector, respectively. Here, $M_{\rm{HWP1,2}}$ are the Jones matrices for the first and second HWPs, and $M_{\rm{refl}}$ is the Jones matrix for reflection from the wire, defined by the reflection coefficients. The Jones matrix for a half-wave plate is
\begin{equation*}
    M_{\rm{HWP}}(\theta_i) = 
        \begin{bmatrix}
        \cos 2\theta_i & \sin 2\theta_i \\
        \sin 2\theta_i & -\cos 2\theta_i
        \end{bmatrix},
\end{equation*}
where $\theta_i$ ($i=$ HWP1, HWP2) is the angle between the $s$-direction and the fast axis of the corresponding HWP. For $\theta_{\rm{HWP1}}=0$ and $\theta_{\rm{HWP2}}=\pi/8$, the balanced signal reduces to $\Delta I = |E_s^{(2)}|^2-|E_p^{(2)}|^2 = -I^{(1)} (r_{ss}r^*_{ps}+r^*_{ss}r_{ps})=-2I^{(1)}|r_{ss}|^2\theta_{s \rightarrow p}$, where $I^{(1)}$ is the beam intensity prior to reflection. We obtained $\theta_{s \rightarrow p}$ by normalizing the signal with $I_0 (= |E_s^{(2)}|^2+|E_p^{(2)}|^2\approx I^{(1)}|r_{ss}|^2)$. Similarly, $\theta_{p \rightarrow s}$ was obtained with $\theta_{\rm{HWP1}}=\pi/4$ and $\theta_{\rm{HWP2}}=\pi/8$.

To measure the ellipticities \(\eta_{s \rightarrow p}\) and \(\eta_{p \rightarrow s}\), a quarter-wave plate (QWP) with its fast axis parallel to the \(s\)-direction was inserted in front of HWP2. The Jones vector before the balanced photo-detector then becomes
\begin{equation*}
    \begin{bmatrix}
        E^{(2)}_s \\
        E^{(2)}_p
    \end{bmatrix}
    =M_{\rm{HWP2}}M_{\rm{QWP}}M_{\rm{HWP1}}M_{\rm{refl}}M_{\rm{HWP1}}
    \begin{bmatrix}
        E^{(1)}_s \\
        E^{(1)}_p
    \end{bmatrix},
\end{equation*}
where the QWP Jones matrix is
\begin{equation*}
    M_{\rm{QWP}}(\theta_j) = \begin{bmatrix}
        \cos^2 \theta_j + i \sin^2 \theta_j & (1-i) \sin \theta_j \cos \theta_j \\
        (1-i) \sin \theta_j \cos \theta_j & \sin^2 \theta_j + i \cos^2 \theta_j
    \end{bmatrix},
\end{equation*}
with $j=$ HWP1, HWP2, QWP.
The ellipticity \(\eta_{s \rightarrow p}\) was obtained at $(\theta_{\rm{HWP1}},\theta_{\rm{HWP2}},\theta_{\rm{QWP}}) = (0,\pi/8,0)$, and \(\eta_{p \rightarrow s}\) at $(\pi/4,\pi/8,0)$.

\subsection*{Extraction of $\sigma_{zx}^{\rm{S}}$ and $\sigma_{zx}^{\rm{A}}$}
From the Fresnel equation, the complex Kerr angles are given by Eq.~(\ref{Eq:KerrFresnel}). 
%Because the proportionality constants $\mathcal{C}_{s\to p}$ and $\mathcal{C}_{p\to s}$ are generally different, simple algebraic combinations of Kerr rotations (such as $\tilde{\theta}_{s\to p} \pm \tilde{\theta}_{p\to s}$) cannot be directly identified with the symmetric and antisymmetric optical conductivity. 
%The two proportionality constants $\mathcal{C}_{s\to p}$ and $\mathcal{C}_{p\to s}$, which can be computed from the incidence angle $\phi_{\rm{i}}$, and the refractive index $n$ then allow $\sigma_{zx}^{\rm{S}(A)}$ and $\tilde{\theta}_{Q(M)}$ to be extracted from the measured Kerr rotations.
The two geometry-dependent constants, $\mathcal{C}_{s\to p}$ and $\mathcal{C}_{p\to s}$, are obtained numerically from the incidence-angle distribution of the focused Gaussian beam. The average incidence angle is $\phi_{\rm i} = \tan^{-1} (\Delta_{\rm BS}/f) = 20.6^\circ$, where $\Delta_{\rm BS}$ is the beam displacement from the optical axis of the objective lens and $f$ is its focal length. Combined with the refractive index $n$, these constants then allow $\sigma_{zx}^{\rm{S}}$ and $\sigma_{zx}^{\rm{A}}$ to be extracted from the measured Kerr angles.

We use literature values of the refractive index $n$ and diagonal optical conductivity $\sigma_{\rm diag}$: $n=1.88+3.32i$ and $\sigma_{\rm diag}=(3.15+2.14i)\times10^5$~S/m for Ti~\cite{Lynch1975}, and $n=2.19+4.40i$ and $\sigma_{\rm diag}=(4.86+3.93i)\times10^5$~S/m for Pt~\cite{Foiles1985}. From the imaginary part of $n$ ($\kappa=3.32$ for Ti and $\kappa=4.40$ for Pt) and the light wavelength $\lambda=660$~nm, we calculate the penetration depth $\ell_{\rm P}=\lambda/(2\pi\kappa)=31.6$~nm for Ti and $\ell_{\rm P}=23.9$~nm for Pt. Note that this $\ell_{\rm P}$ is the electric-field-amplitude penetration depth and is twice the intensity penetration depth.
%The refractive index $n$ was determined using the same optical setup. The intensity difference measured by the balanced photo-detector is given by
%\begin{equation}
%    \begin{split}
%        \Delta I (0,\theta_{\rm{HWP2}}) &= I^{(1)}|r_{ss}|^2 \cos (4 \theta_{\rm{HWP2}}), \\
%        \Delta I (\pi/4,\theta_{\rm{HWP2}}) &= I^{(1)}|r_{pp}|^2 \cos (4 \theta_{\rm{HWP2}}), \\
%        \frac{\Delta I (3\pi/8,\theta_{\rm{HWP2}}) + \Delta I (\pi/8,\theta_{\rm{HWP2}})}{2} &= + I^{(1)} |r_{ss}||r_{pp}| \cos (4\theta_{\rm{HWP2}}) \cos \Delta, \\
%        \frac{\Delta I (3\pi/8,\theta_{\rm{HWP2}}) - \Delta I (\pi/8,\theta_{\rm{HWP2}})}{2} &= - I^{(1)} |r_{ss}||r_{pp}| \sin (4\theta_{\rm{HWP2}}) \sin \Delta,
%    \end{split}
%    \label{Eq:noCurrent}
%\end{equation}
%where $\Delta = \arg (r_{pp}/r_{ss})$. The $\pi/2$-periodic balanced photo-detector signals are shown in fig.~\ref{Fig:S4}. Combining the measured $r_{pp}/r_{ss}$ with the known incidence angle $\phi_{\rm{i}}$, the refractive index $n$ is uniquely determined from the Fresnel equations. Because each Ti film is covered by a capping layer that introduces multiple reflections, the extracted $n$ is an effective refractive index; for our samples we find $n_{\rm{Ti}}=2.22-i0.68$, from which the diagonal optical conductivity is straightforwardely obtained [$\sigma_{\rm diag}= (7.63+i8.75)\times10^4$~S/m].

\subsection*{First-principles calculation of conversion factors between multipole density and optical conductivity}

The optical-conductivity responses associated with spin, OAM, orbital-quadrupole, and electric-quadrupole accumulations were calculated for bulk hexagonal close-packed (hcp) Ti and bulk face-centered cubic (fcc) Pt. A fictitious Zeeman-like interaction was introduced to generate a uniform accumulation. Following the approach employed in previous magneto-optical calculations~\cite{Choi2023,Stamm2017}, the perturbation Hamiltonian corresponding to the Zeeman-like interaction is defined as
\begin{equation}
H_Z=J\mathcal{O},
\label{Eq:Zeemanlike}
\end{equation}
where $\mathcal{O}=S_i,\,L_i,\,\{L_i,L_j\},\,Q_{ij}$ for spin, OAM, orbital quadrupole, and electric quadrupole, respectively ($i,j\in\{x,y,z\}$). %Here, \(J\) is expressed in units of \(\mathrm{eV}/\hbar\), \(\mathrm{eV}/\hbar\), \(\mathrm{eV}/\hbar^2\), and \(\mathrm{eV}/Q_{\mathrm{unit}}\) for the spin, OAM, OQ, and EQ, respectively. %The reference EQ unit was defined as \(Q_{\mathrm{unit}}^{\mathrm{Ti}}=-e a_{\mathrm{Ti}}c_{\mathrm{Ti}}/4\) for hcp Ti, where \(a_{\mathrm{Ti}}\) and \(c_{\mathrm{Ti}}\) are its lattice constants, and as \(Q_{\mathrm{unit}}^{\mathrm{Pt}}=-e a_{\mathrm{Pt}}^2/4\) for fcc Pt, where \(a_{\mathrm{Pt}}\) is its lattice constant.

The density functional theory calculations were performed using the OpenMX package~\cite{Ozaki2003}. Bulk hcp Ti was described using lattice constants $a=0.295$~nm and
$c=0.468$~nm and an \(s2p2d3\) PAO basis set. Bulk fcc Pt was described using a lattice constant $b=0.392$~nm and an \(s2p1d2\) basis set. Then the Zeeman-like interaction [Eq.~(\ref{Eq:Zeemanlike})] was added to the Hamiltonian, and the optical conductivity was calculated within linear-response theory~\cite{Wang1974,Kunes1999}:
\[
\begin{aligned}
\sigma_{\beta\alpha}(\omega)
={}&
\frac{i e^2\hbar}{\Omega_{\mathrm{cell}}}
\sum_{\boldsymbol{k}}\sum_{jl}
\frac{f_0(\epsilon_{l\boldsymbol{k}})
\left[1-f_0(\epsilon_{j\boldsymbol{k}})\right]}{\epsilon_{j\boldsymbol{k}}-\epsilon_{l\boldsymbol{k}}}
\left[
\frac{\mathcal{V}_{lj}^{\beta}(\boldsymbol{k})\mathcal{V}_{jl}^{\alpha}(\boldsymbol{k})}
{\hbar\omega+\epsilon_{l\boldsymbol{k}}-\epsilon_{j\boldsymbol{k}}+i\delta}+
\frac{\mathcal{V}_{lj}^{\beta *}(\boldsymbol{k})\mathcal{V}_{jl}^{\alpha *}(\boldsymbol{k})}
{\hbar\omega+\epsilon_{j\boldsymbol{k}}-\epsilon_{l\boldsymbol{k}}+i\delta}
\right],
\end{aligned}
\]
where \(\Omega_{\mathrm{cell}}\) is the unit-cell volume,
\(f_0(\epsilon_{n\boldsymbol{k}})\) is the Fermi--Dirac distribution,
\(\epsilon_{n\boldsymbol{k}}\) and
\(|u_{n\boldsymbol{k}}\rangle\) are the energy eigenvalue and eigenstate
of band \(n\), respectively, and
\(\mathcal{V}_{jl}^{\alpha}(\boldsymbol{k})
=\langle u_{j\boldsymbol{k}}|v_{\alpha}|u_{l\boldsymbol{k}}\rangle\)
is the corresponding matrix element of the velocity operator \(v_{\alpha}\) along the \(\alpha\) direction.
A level broadening of \(\delta=0.4~\mathrm{eV}\) was used, and the optical response was evaluated at a photon energy of \(\hbar\omega=1.88~\mathrm{eV}\), corresponding to the 660-nm laser wavelength used in the experiment.

For each perturbation, the induced expectation value of the corresponding spin, OAM, orbital quadrupole, or electric quadrupole operator was calculated from the perturbed eigenstates as
\(\Delta \mathcal{O}=\left\langle\mathcal{O}\right\rangle_J-\left\langle\mathcal{O}\right\rangle_{J=0}.\)
The optical-conductivity response per unit induced spin, OAM, orbital quadrupole, or electric quadrupole was then evaluated as
\(
\sigma_{\beta\alpha}^{\mathcal{O},\mathrm{unit}}(\omega)
=(\sigma_{\beta\alpha}(\omega;J)-\sigma_{\beta\alpha}(\omega;0))/\Delta\mathcal{O}.
\)
Because the experimental films are polycrystalline, we compute $\sigma_{\beta\alpha}^{\mathcal{O},\mathrm{unit}}(\omega)$ for various orientations. 
The orientation-averaged $\sigma_{\beta\alpha}^{\mathcal{O},\mathrm{unit}}(\omega)$ is used to convert the measured $\sigma_{\rm sat}^{\rm S}$ and $\sigma_{\rm sat}^{\rm A}$ into associated multipole accumulations.

%%%%%%%%%%%%%%%%%%%%%%%%%%%%%%%%%%%%%%%%%%%%%%%%%%%%%%%%%%%%%%%
\newpage
\def\bibsection{\section*{Methods references}}

\newpage

\newpage
\setcounter{figure}{0}
\renewcommand{\thefigure}{\arabic{figure}}
\renewcommand{\figurename}{\textbf{Extended Data Fig.}}

%\begin{figure*}[h]
%\centering
%\includegraphics[width=10cm]{SFigure5_260521.pdf}
%\caption{\justifying \textbf{Thickness dependence of the electrical resistivity.} Resistivity $\rho$ of the Ti thin films as a function of the Ti thickness $t_{\mathrm{Ti}}$. The resistivity remains approximately $3~\mu\Omega \cdot \mathrm{m}$ for $t_{\mathrm{Ti}} \geq 20$~nm, with a slight enhancement at $t_{\mathrm{Ti}} = 10$~nm attributed to interfacial scattering.}
%\label{Fig:S5}
%\end{figure*}

\begin{figure*}[h]
	\centering
    \includegraphics[width=10.5cm]{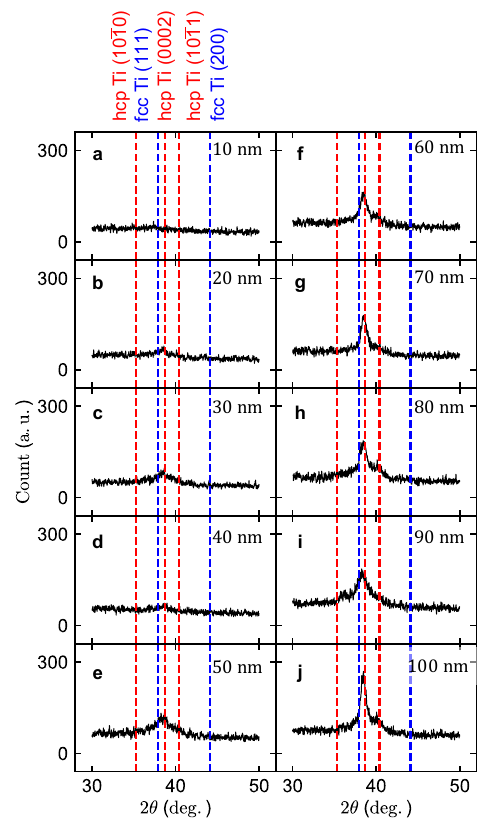}
    \caption{\justifying \textbf{X-ray diffraction of Ti thin films.} \(2\theta\) scans of Ti thin films on \(c\)-cut sapphire substrates for thicknesses ranging from 10~nm to 100~nm (\textbf{a--j}). Black solid lines show the experimental data, and red (blue) dashed lines mark $2\theta$-peak positions of the hexagonal-close-packed (face-centered-cubic) structure. The dominant hcp $(0002)$ peaks, and shallow $(10\bar{1}0)$ and $(10\bar{1}1)$ peaks are observed, with negligible fcc $(111)$ and $(200)$ contributions, confirming the hcp structure of the deposited Ti films.
\label{Fig:S1} }
\end{figure*}

\begin{figure*}[h]
	\centering
    \includegraphics[width=10.5cm]{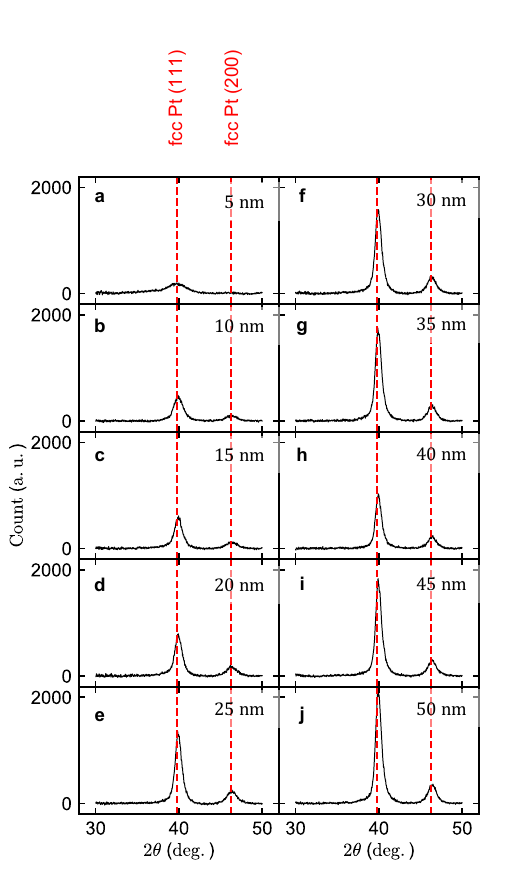}
    \caption{\justifying \textbf{X-ray diffraction of Pt thin films.} \(2\theta\) scans of Pt thin films on \(c\)-cut sapphire substrates for thicknesses ranging from 5~nm to 50~nm (\textbf{a--j}). Black solid lines show the experimental data, and red dashed lines mark $2\theta$-peak positions of the face-centered-cubic structure. The dominant fcc $(111)$, and $(200)$ peaks are observed.
\label{Fig:SPt} }
\end{figure*}

\begin{figure*}[h]
\centering
\includegraphics[width=10cm]{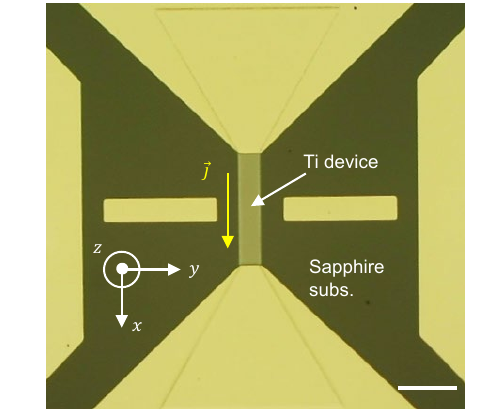}
\caption{\justifying \textbf{Optical microscope image of the Ti device.} The central narrow region is the Ti wire (the device under measurement), patterned on a $c$-cut sapphire substrate and contacted by Au electrode pads on both ends. The yellow arrow indicates the direction of the applied current density $j$, and the coordinate system $(x,y,z)$ defines the measurement geometry, with $z$ along the surface normal. Scale bar: 50~{\textmu}m.}
\label{Fig:S2}
\end{figure*}

\begin{figure*}[h]
\centering
\includegraphics[width=8.5cm]{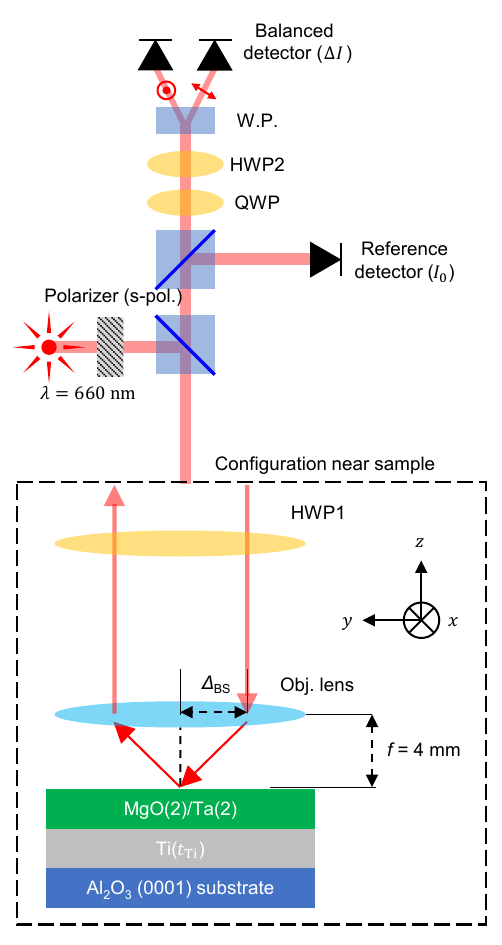}
\caption{\justifying \textbf{Schematic illustration of the optical setup.} A 660 nm continuous-wave laser passes through a polarizer to produce $s$-polarized light, which is then directed by a beam splitter toward the sample. The dashed box highlights the configuration near the sample: a half-wave plate (HWP1) and an objective lens ($f = 4$~mm) focus the light onto the Ti film. The incidence angle is controlled by the lateral displacement $\Delta_{\mathrm{BS}}$ of the first beam splitter. After reflection, the beam is split into two paths by a second beam splitter: one reaches a reference detector measuring the reflected intensity $I_0$, and the other passes through a quarter-wave plate (QWP), a second half-wave plate (HWP2), and a Wollaston prism (W.P.) before being detected by a balanced detector ($\Delta I$). The QWP is inserted only for ellipticity measurements. Numbers in parentheses indicate layer thicknesses in nanometers.}
\label{Fig:S3}
\end{figure*}

\end{document}